# Astrophysicists on Twitter:
## An in-depth analysis of tweeting and scientific publication behavior

Stefanie Haustein[*,1], Timothy D. Bowman[2], Kim Holmberg[3], Isabella Peters[4] & Vincent Larivière[1,5]

[*]stefanie.haustein@umontreal.ca

[1] École de bibliothéconomie et des sciences de l'information (EBSI), Université de Montréal, Montréal, QC (Canada)

[2] School of Informatics and Computing, Indiana University, Bloomington, IN (USA)

[3] School of Technology, University of Wolverhampton, Wolverhampton (UK)

[4] ZBW Leibniz Information Centre for Economics & Christian Albrechts University Kiel, Kiel (Germany)

[5] Observatoire des Sciences et des Technologies (OST), Centre Interuniversitaire de Recherche sur la Science et la Technologie (CIRST), Université du Québec à Montréal, Montréal, QC (Canada)

**Abstract**

**Purpose** – This paper analyzes the tweeting behavior of 37 astrophysicists on Twitter and compares their tweeting behavior with their publication behavior and citation impact to show whether they tweet research-related topics or not.

**Design/methodology/approach** – Astrophysicists on Twitter are selected to compare their tweets with their publications from Web of Science. Different user groups are identified based on tweeting and publication frequency.

**Findings** – A moderate negative correlation ($\rho=-0.390$*) is found between the number of publications and tweets per day, while retweet and citation rates do not correlate. The similarity between tweets and abstracts is very low ($cos=0.081$). User groups show different tweeting behavior such as retweeting and including hashtags, usernames and URLs.

**Research limitations/implications** – The study is limited in terms of the small set of astrophysicists. Results are not necessarily representative of the entire astrophysicist community on Twitter and they most certainly do not apply to scientists in general. Future research should apply the methods to a larger set of researchers and other scientific disciplines.

**Practical implications** – To a certain extent, this study helps to understand how researchers use Twitter. The results hint at the fact that impact on Twitter can neither be equated with nor replace traditional research impact metrics. However, tweets and other so-called altmetrics might be able to reflect other impact of scientists such as public outreach and science communication.

**Originality/value** – To the best of our knowledge, this is the first in-depth study comparing researchers' tweeting activity and behavior with scientific publication output in terms of quantity, content and impact.

**Article Classification**: Research paper

## 1 Introduction

Over the last few years, the use of Twitter and other social media has widespread in the various spheres of society, including the scientific community. Some scientists use social media tools for collaborative authoring, scheduling meetings, conferencing, disseminating research results and discovering new information and research ideas (Rowlands *et al.*, 2011). This has led to the development of a new family of scientific indicators based on the mentions of scientific output on general social media platforms, such as Twitter, Facebook or LinkedIn, and more research-focused services such as Mendeley, CiteULike and ResearchGate etc. Despite their already widespread use (and the hype created around them), very little is known about the extent to which social media are used in scholarly communication, especially in regards to how and why (Kieslinger *et al.*, 2011), as well as their relationship with traditional indicators of research impact. Earlier research has focused on the use of Twitter at scientific conferences and shown how Twitter is used as a backchannel at conferences and as a tool to reach wider audiences (Letierce *et al.*, 2010; Reinhardt *et al.*, 2009; Weller and Puschmann, 2011; Weller *et al.*, 2011). Surveys demonstrated that the use of Twitter and other micro-blogging platforms among researchers ranges between 7.3% (Rowlands *et al.*, 2011) and 17.7% (Ponte and Simon, 2011). Eysenbach (2011) and Shuai *et al.* (2012) have found a connection between the number of tweets about articles and later citation counts. In biomedicine, Haustein *et al.* (2014) and Thelwall *et al.* (2013) have shown on a large scale that, although correlated, social media mentions and citations were not identical, and were actually indicators of a different kind of impact. However, results may differ between journals and specialties, as there are disciplinary differences in researchers' tweeting behavior (Holmberg and Thelwall, 2013; Haustein *et al.*, 2014).

The goal of this paper is to contribute to the discussion of the meaning of so-called altmetrics, by analyzing the tweeting behavior of a sample of 37 astrophysicists active on Twitter (Holmberg, 2013) in depth and compare it to their publication output and citation impact in scholarly journals. The following research questions are addressed:
1. How do astrophysicists use Twitter? How often do they tweet and which of the Twitter affordances (retweets, hashtags, @messages) do they use?
2. How does tweeting behavior compare to publication behavior? How do research output and impact (citations) compare to Twitter activity and impact (followers, retweets)?
3. What do astrophysicists tweet about? Are the contents of their tweets similar to their scientific papers?

The results will provide evidence in how far selected astrophysicists use Twitter for scholarly communication and if their activity and impact on Twitter and in scientific journals are related.

## 2 Methods
*2.1 Data collection*
A previous study (Holmberg & Thelwall, 2013) identified astrophysicists on Twitter by searching for astrophysics-related keywords and checking personal Twitter and Web profiles of users. Although all Twitter users in our sample have a background in astrophysics, the 37 selected Twitter users represent a mix of established researchers, professors, astrophysicists working in observatories and science communicators. We chose astrophysicists for our analysis since they form a coherent community around a specific enough topic which is also, in terms of language, better distinguishable from other, more general topics in tweets (e.g., when compared to economics). We also assumed that the popularity of this scientific field among the general public would attract more researchers interested in publicly communicating scientific findings (Bauer, 2013). In order to analyze how Twitter is used by these astrophysicists, we analyzed tweets in depth and compared the results to the researchers' publication output and citation rates. Information such as full name and affiliated institutions of 37 astrophysicists on Twitter were identified so that their publication records could be retrieved from Web of Science (WoS). Publications were extracted from the Science Citation Index in April 2013 covering all documents published during the 5-year period from 2008 to 2012. Author names were cleaned manually to ensure validity of publication records. The citation window covers all citations received until the end of 2012.

Tweets were collected in May 2013 and covered all available tweets per user name. As the Twitter API restricts results to the approximately 3,200 most recent tweets, the entire tweeting history could not be collected for those astrophysicists that had published more than 3,200 tweets. However, the total number is indicated on the user profiles and was thus recorded to obtain an indicator of the complete tweeting activity. Overall, the 37 astrophysicists published 289,368 tweets of which 68,232 could be retrieved. While the comparison of publication and Twitter activity is based on the total number of tweets, the structural and content analysis had to be restricted to the 68,232 tweets available for download. The actual number of tweets was considerably higher just for 12 of the 37 users so that for 25 users the complete tweeting history was considered for detailed analysis. In addition, we also collected the number of followers and the date the users opened their Twitter accounts. Correlations between various publication- and Twitter-based indicators were computed using Spearman's $\rho$.

*2.2 Determining user groups*
Astrophysicists were grouped into different user groups according to their tweeting and publication behavior. Tweeting characteristics were compared across these groups in order to determine whether, for example, active Twitter users retweet more frequently or if those who publish frequently distribute more URLs. Publication behavior was defined by the number of papers published in WoS journals during the five-year period 2008 to 2012 and tweeting behavior is based on the average number of tweets per day. Groups were determined by choosing four levels of publication and tweeting frequency, respectively. In terms of publication behavior, we distinguished between astrophysicists who *do not publish* (0 publications; 6 astrophysicists), *publish occasionally* (1-9; 13), *publish regularly* (14-37; 14) and those who *publish frequently* (46-112; 5). In order to account for the different tweeting windows the number of total tweets was divided by the number of days since the user created his or her Twitter account [1]. There were no correlations between the number of days since signing up and the number of total tweets or followers, indicating that, at least for our dataset, age on Twitter is not correlated with activity or with indirect popularity. The astrophysicists were divided into those who *tweet rarely* (0.0 to 0.1 tweets per day; 5 astrophysicists), *tweet occasionally* (0.1-0.9; 11), *tweet regularly* (1.2-2.9; 11) and *tweet frequently* (3.7-58.2; 10).

**Table 1.** Number of astrophysicists assigned to groups reflecting their level of tweeting and publication activity.

| Selected astrophysicists *(N=37)* | tweet rarely *(0.0-0.1 tweets per day)* | tweet occasionally *(0.1-0.9)* | tweet regularly *(1.2-2.9)* | tweet frequently *(3.7-58.2)* | total (publishing activity) |
|---|---|---|---|---|---|
| do not publish *(0 publications 2008-2012)* | -- | -- | 1 | 5 | **6** |
| publish occasionally *(1-9)* | 4 | 3 | 4 | 2 | **13** |
| publish regularly *(14-37)* | -- | 5 | 5 | 3 | **13** |
| publish frequently *(46-112)* | 1 | 3 | 1 | -- | **5** |
| total (tweeting activity) | **5** | **11** | **11** | **10** | **37** |

Table 1 presents the number of astrophysicists assigned to the 16 groups reflecting both their level of tweeting and publication activity. Among the 37 researchers there were none that do not publish and tweet only rarely or occasionally, nor were there any that tweet and publish frequently. In addition, none of the 37 astrophysicists tweeted rarely and published regularly. Hence, out of the 16 possible groups only 12 actually exist. Tweeting characteristics such as the number of retweets and tweets containing hashtags, user names and URLs were calculated for each of the 12 groups to investigate whether users classified according to their publication and tweet frequency use Twitter differently. In order to weigh the tweeting behavior of each astrophysicist equally and to normalize for different group sizes and number of tweets per person, group values are calculated as means of percentages of tweets with a particular characteristic compared to all tweets per person.

*2.3 Tweeting characteristics of user groups*
The 68,232 tweets downloaded via the Twitter API were analyzed from a structural perspective identifying the number of tweets containing user names (identified by @username), hashtags (identified by # followed by a string of characters up to a blank), URLs (identified by a string of characters starting with http://), the number of retweets and the number of times each tweet was itself retweeted. While the former type of retweets could be compared to the number of papers an author cites (references), the latter could be comparable to the number of citations an author receives. Weller and Puschmann (2011) refer to these as internal and external citations. Retweets were limited to "plain retweets" (Burgess and Bruns, 2012) that begin with the string "RT @user name". Although other methods indicating forwarded tweets are possible (e.g., "via @user name" and "MT" for modified tweet), we restricted the analysis to the so-called "plain retweets". The number of times a user's tweet has itself been retweeted (the tweet citations), as indicated in the tweet metadata, is determined the same way. In our dataset plain retweets make up 91.3% of all retweets.

*2.4 Comparison of tweet and publication contents*
In order to find out whether astrophysicists tweet and publish about similar topics, tweet terms were compared to terms found in abstracts. Hashtags were included and treated as ordinary words. This part of the analysis was reduced to those authors who publish at least regularly, that is a minimum of 14 publications during the 2008 to 2012 period, to ensure the availability of a significant amount of abstract terms. Seventeen out of 18 authors published papers in 2008, whereas one author has no publications in 2008 but published several articles between 2009 and 2012. Ten authors started tweeting in 2007, 2008, or 2009 and seven authors tweeted in 2010 or 2011 for the first time. For each of those 18 authors, noun phrases were extracted from tweets and abstracts with the linguistic filter of the natural language processing tool provided by VOSviewer (Van Eck *et al.*, 2010), which is based on the part-of-speech tagger developed by Schmid (1994). The two respective term sets were compared using an exact match of character strings for each of the 18 authors and the combined term sets of all authors. Similarities were computed using the cosine similarity measure (Salton and McGill, 1987). As the combined set of terms contained 50,854 different noun phrases in the tweets and 12,970 in the abstracts, the analysis was confined to the automatic method of matching entire character strings, although this reduces the overlap because synonyms are not merged. Note, however, that the number of term variants is limited when only noun phrases are considered (as opposed to other parts of speech) and that VOSviewer already merges regular singular and plural forms (Van Eck *et al.*, 2010). Stemming, as applied by Haustein and Peters (2012), was thus not necessary. Although we acknowledge that the use of a proper astrophysics thesaurus (e.g., which would assist in uniting, for example, *CME* and *coronal mass ejection*, Figure 4) would have been more accurate, this was not feasible in our study.

Nevertheless, we assume that this limitation applies to all tweets and abstracts of the users and thus the similarity values we provide rather underestimate actual similarities between tweet and abstract terms (Haustein and Peters, 2012). Given that there are different publication windows for tweets and articles, where in one case a paper was published in 2008 but the first tweet was sent in 2012, the vocabulary in tweets and abstracts may vary because of changed research foci. On the other hand, tweets are timelier and can be sent before the articles were published. Hence, we forgo the year-wise normalization of term matches in favor of analyzing the bag of terms used in publications and tweets.

## 3 Results and Discussion
### 3.1 Comparing overall tweeting and publication behavior
The mean number of total tweets per astrophysicists in our dataset was 7,821, although data was heavily skewed, ranging from 27 tweets to 90,835 around a median of 2,275. On average, the number of papers in WoS published during the 2008 to 2012 period was 17.6 per researcher (median: 9); the most productive researcher authored 112 documents and 6 astrophysicists did not publish at all. As shown in Figure 1 and by a Spearman correlation of -0.339** between number of publications and tweets per day (Table 2), there is a moderate negative correlation between publication and Twitter activity, indicating that in our sample those who publish more, tweet less, and vice versa.

Two astrophysicists stand out in terms of tweeting behavior, with almost 60 tweets per day. Both of them have not published any journal articles during the 5-year period analyzed. A closer investigation of these astrophysicists indicates that one is an aspiring science communicator and the other works as a science communicator and journalist. The researcher who published most actively among the selected scientists authored 112 documents and tweeted only occasionally (421 tweets since June 2008, 0.23 tweets per day). Among the 10 most frequent tweeters (more than 3 tweets a day) there are 3 that publish regularly (11.15 tweets per day/19 publications; 5.01/27; 4.03/37), indicating that there are also some very active researchers who frequently use Twitter. Whether this is for communicating research or for private reasons will be investigated below by comparing tweet and abstract terms.

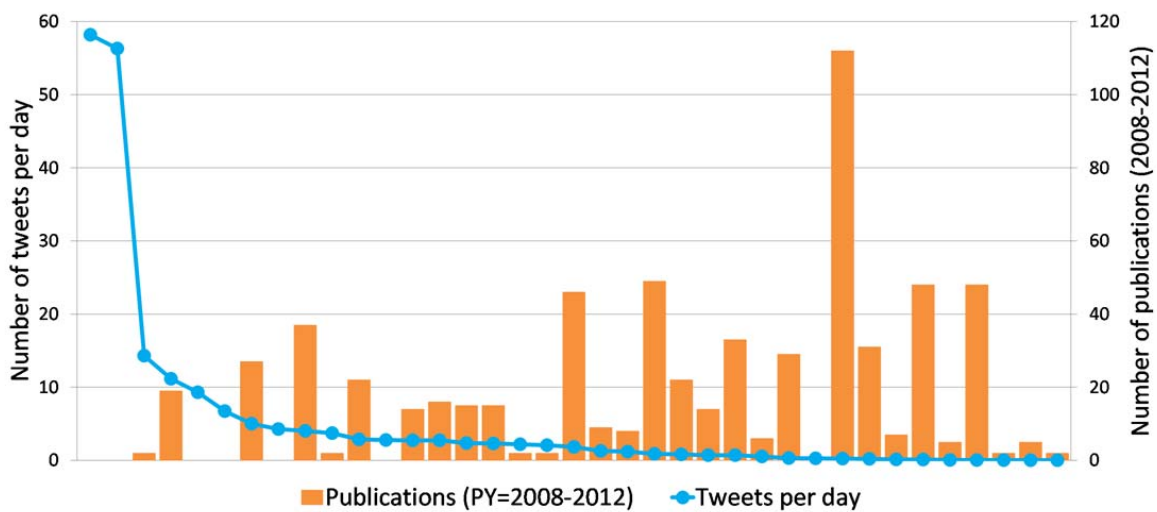

**Figure 1.** Number of tweets normalized by time since joining Twitter and number of documents published in WoS journals between 2008 and 2012 per astrophysicist.

There is a negative relation between scientific impact and Twitter activity: the higher the citation rate, the less tweets (-0.431**) and followers (-0.350*). This implies that, among those astrophysicists analyzed, those that are highly active and popular on Twitter are not high impact researchers in terms of publications and citations. Investigations of Twitter profiles and homepages suggest that many of the very active tweeters have often taken on the role of science communicators and public outreach officers. No correlation can be observed between the citation and retweet rate (0.077) and the number of citations and retweets (-0.040), respectively. As indicated by $\rho=0.500**$ between the number of tweets and followers, there is a medium positive correlation between the level of activity and the size of the direct audience. However, a large amount of tweets does not automatically guarantee many followers and vice versa. In fact, a well-known scientist in our dataset had published less than 50 tweets, yet he had almost 200,000 followers on Twitter.

**Table 2.** Spearman's $\rho$ comparing publication and Twitter indicators for the 37 selected astrophysicists on Twitter ($P$: number of documents in WoS published between 2008 and 2012; $C$: sum of citations to $P$; $C/P$: average citation rate; $T_{total}$: number of total tweets as indicated on the Twitter profile; $F$: number of Twitter followers; *active*: time since Twitter sign up in days; $T_{total}/active$: average number of tweets per day; $T_{col}$: number of tweets collected; $RT_{col}$: number of times collected tweets were retweeted; $RT_{col}/T_{col}$: retweet rate).

| $\rho$ | $P$ | $C$ | $C/P$ | $T_{total}$ | $F$ | active | $T_{total}/active$ | $T_{col}$ | $RT_{col}$ | $RT_{col}/T_{col}$ |
|---|---|---|---|---|---|---|---|---|---|---|
| $P$ | | .939** | .635** | -.360* | -.196 | -.053 | -.339* | -.278 | -.082 | .023 |
| $C$ | .939** | | .826** | -.368* | -.245 | .038 | -.375* | -.299 | -.040 | .057 |
| $CPP$ | .635** | .826** | | -.431** | -.350* | .077 | -.457** | -.381* | -.069 | .077 |
| $T_{total}$ | -.360* | -.368* | -.431** | | .500** | 0.096 | .965** | .959** | .612** | .183 |
| $F$ | -.196 | -.245 | -.350* | .500** | | .147 | .482** | .495** | .277 | .113 |
| active | -.053 | .038 | .077 | .096 | .147 | | -.104 | .041 | .203 | .147 |
| $T_{total}/active$ | -.339* | -.375* | -.457** | .965** | .482** | -.104 | | .944** | .541** | .122 |
| $T_{col}$ | -.278 | -.299 | -.381* | .959** | .495** | .041 | .944** | | .611** | .191 |
| $RT_{col}$ | -.082 | -.040 | -.069 | .612** | .277 | .203 | .541** | .611** | | .842** |
| $RT_{col}/T_{col}$ | .023 | .057 | .077 | .183 | .113 | .147 | .122 | .191 | .842** | |

* correlations are significant at the 0.05 level, two-tailed. ** correlations are significant at the 0.01 level, two-tailed.

*3.2 Tweeting characteristics of user groups*

Although there is a general trend that the more astrophysicists publish, the less they tweet, Figure 1 and the moderate negative correlation between number of publications and tweets per day show that there are various kinds of users. To discover overall differences in how astrophysicists use Twitter, the extent to which they retweet, send URLs, use hashtags and address other users in their tweets, was compared among the 12 groups defined above.

**Table 3.** Mean of share of retweets per person per group.

| Retweets | tweet rarely | tweet occasionally | tweet regularly | tweet frequently | total (publishing activity) |
|---|---|---|---|---|---|
| do not publish | -- | -- | 23.8% | 11.9% | **13.9%** |
| publish occasionally | 5.5% | 6.7% | 12.7% | 35.4% | **12.6%** |
| publish regularly | -- | 14.4% | 9.6% | 11.7% | **11.9%** |
| publish frequently | 0.0% | 25.4% | 7.2% | -- | **16.7%** |
| total (tweeting activity) | **4.4%** | **15.3%** | **11.8%** | **16.5%** | **13.1%** |

Of the 68,232 tweets analyzed, 9,914 could be identified as plain retweets, which constitutes to a share of 14.5% and an average of 13.1% per person (Table 3). Those astrophysicists that tweet rarely, hardly use the retweet function; on average, only 4.4% of their tweets are retweets. Frequent tweeters retweet more than average (16.5%) and among those occasional authors retweet often. More than one third (35.4%) of the tweets are retweets among those who published between one and nine articles and tweet more than three times a day. Those astrophysicists who publish frequently, that is between 46 and 112 papers during the period analyzed, and tweet occasionally (0.1 to 0.9 tweets per day), retweet almost twice as much as average.

**Table 4.** Mean of share of tweets that contain at least one hashtag per person per group.

| Hashtags | tweet rarely | tweet occasionally | tweet regularly | tweet frequently | total (publishing activity) |
|---|---|---|---|---|---|
| do not publish | -- | -- | 17.7% | 20.2% | **19.8%** |
| publish occasionally | 1.2% | 8.3% | 26.4% | 15.3% | **12.8%** |
| publish regularly | -- | 43.6% | 17.2% | 36.6% | **31.8%** |
| publish frequently | 5.5% | 39.3% | 20.9% | -- | **28.9%** |
| total (tweeting activity) | **2.1%** | **32.8%** | **20.9%** | **24.1%** | **22.8%** |

**Table 5.** Mean of share of tweets that contain at least one user name per person per group.

| User names | tweet rarely | tweet occasionally | tweet regularly | tweet frequently | total (publishing activity) |
|---|---|---|---|---|---|
| do not publish | -- | -- | 85.3% | 62.8% | **66.6%** |
| publish occasionally | 20.9% | 37.0% | 45.4% | 54.4% | **37.3%** |
| publish regularly | -- | 53.6% | 70.1% | 66.6% | **62.9%** |
| publish frequently | 2.7% | 52.7% | 44.7% | -- | **41.1%** |
| total (tweeting activity) | **17.3%** | **48.8%** | **60.2%** | **62.3%** | **51.6%** |

Compared to the retweet function, the use of hashtags seems to be more common among the astrophysicists. Almost one quarter of the examined tweets contained at least one hashtag (23.4%). By contrast, boyd et al. (2010) showed that 5% of general tweets contained hashtags. However, large differences can be observed among user groups. The scientists that rarely tweet, hardly make use of that function; only 6 of the 259 tweets published by the group of astrophysicists that tweet rarely, contain a hashtag. Occasional tweeters who publish regularly or who publish frequently and frequent tweeters who publish regularly use hashtags most, as on average 43.6%, 39.3% and 36.6% of their tweets contain hashtags (Table 4). Overall, the 68,232 tweets contained 4,301 unique hashtags, which varied according to the frequency of use and users. It should be noted that the most frequent hashtag #fb (used 546 times) does not represent a hashtag in the original sense but carries out a particular function. When added to the end of the tweet, it triggers the Selective Tweets [2] application on Facebook to add the content of the tweet as a status on Facebook and hence it has more to do with functionality than tweet content. This can be explained by the very active Astronomers Facebook group, which has currently more than

7,500 members. The remaining 9 of the Top 10 hashtags include #sftc [=UK Science & Technology Facilities Council] (410 times), #astrofact (349 times), #jwst [=James Webb Space Telescope] (339 times), #math (332 times), #twinkletweet [=hashtag used by some astrophysicists to label astronomical content] (284 times), #nasa (257 times), #scipolicy (216 times), #aas218 [=American Astrological Society 218[th] Meeting] (194), and #hubble (164 times).

Of the 68,232 tweets, 60.9% contained at least one other user name —46.4% excluding pure retweets— making it the most frequently used among the analyzed Twitter affordances. This shows that Twitter is often used to address others directly, indicating personal communication or discussion, or at least to mention them. The mean of average shares of the 37 users amounts to 51.6% (Table 5) or 38.4% excluding pure retweets. A clear trend can be observed that the more people tweet, the higher the share of tweets that contain user names. While, on average, as few as 17.3% of tweets sent by scientists who tweet rarely include one, the share of tweets with user names increases to 48.8% for occasional and 60.2% and 62.3% for regular and frequent tweeters, respectively (Table 5). Among the 12 groups of astrophysicists, the regular tweeters who do not publish send the most personal tweets (85.3%), while the rarely tweeting astrophysicists who publish the most hardly address other Twitter users (2.7%). Frequent tweeters that do not publish (62.8%) or publish regularly (66.6%) and regular tweeters who publish regularly (70.1%) also include user names well above average.

Table 6. Mean of share of tweets that contain at least one URL per person per group.

| URLs | tweet rarely | tweet occasionally | tweet regularly | tweet frequently | total (publishing activity) |
|---|---|---|---|---|---|
| do not publish | -- | -- | 32.7% | 30.9% | **31.2%** |
| publish occasionally | 45.3% | 23.9% | 24.7% | 68.7% | **37.6%** |
| publish regularly | -- | 33.5% | 27.5% | 53.8% | **35.9%** |
| publish frequently | 30.1% | 46.5% | 45.1% | -- | **42.9%** |
| total (tweeting activity) | **42.3%** | **34.4%** | **28.6%** | **45.3%** | **36.7%** |

A little more than one third of tweets contains URLs. This ratio is comparable to results of other studies in scientific contexts finding 12-43% of conference-related tweets with URLs (Mahrt *et al.*, 2014) and 55% in a group of tweeting scholars of different disciplines (Weller and Puschmann, 2011). In a random sample of general tweets, the use of URLs is slightly lower (20%, boyd *et al.*, 2010). Although differences can be observed between groups, no clear trend emerges regarding publication or tweet frequency (Table 6). Moreover, the group values do not deviate from the mean of 36.7% as much as observed for the other Twitter affordances. Especially the two groups of occasionally (45.3%) and frequently (30.1%) publishing scientists who rarely tweet include URLs much more often than they make use of the Twitter affordances described above. While the highest share of tweets with URLs are sent by frequent tweeters who authored at least one WoS publication (publish occasionally: 68.7%; publish regularly: 53.8%), authors with more than 45 papers who, on average, tweet at least once in 10 days also send more URLs than average (tweet occasionally: 46.5%; tweet regularly: 45.1%).

*3.3 Comparison of tweet and publication content*

To find out whether astrophysicists that are active researchers also use Twitter to communicate research results either to their peers or an interested public (see Haustein *et al.* (2014) for a theoretical framework of Twitter audiences of scholarly papers) tweet terms were compared to abstract terms. For the 18 most frequently publishing astrophysicists a total of 50,854 unique noun phrases were extracted from the dataset of 31,458 tweets (Table 7). The abstracts of the 597 articles contained 12,970 unique noun phrases. Overall, with a cosine of 0.081, the overlap of the two combined term sets was very low, indicating that the selected astrophysicists do not tweet about the same topics as they publish or that they use different vocabularies. As few as 2,075 noun phrases were used in both tweets and abstracts, which amounts to 4.1% of noun phrases used on Twitter and 16.0% in abstracts of scholarly documents. The most frequently used noun phrases in the abstracts of the scholarly literature were *galaxy* (203 times in abstracts; 147 times in tweets), *results* (192; 116), *star* (168; 240), *cluster* (152; 41), *data* (135; 143), *M-circle dot* [=solar mass] (122; 0), *observation* (118; 31), *system* (117; 40), *kpc* [=kiloparsec] (109; 8) and *sample* (104; 13). On the other hand, the terms that occurred most often in the tweets were *today* (690 times in tweets; 2 times in abstracts), *thank* (676; 0), *day* (573; 43), *time* (504; 54), *year* (498; 41), *science* (446; 3), *sun* (378; 15), *NASA* (369; 1), *person* (362; 0) and *twinkletweet* (335; 0).

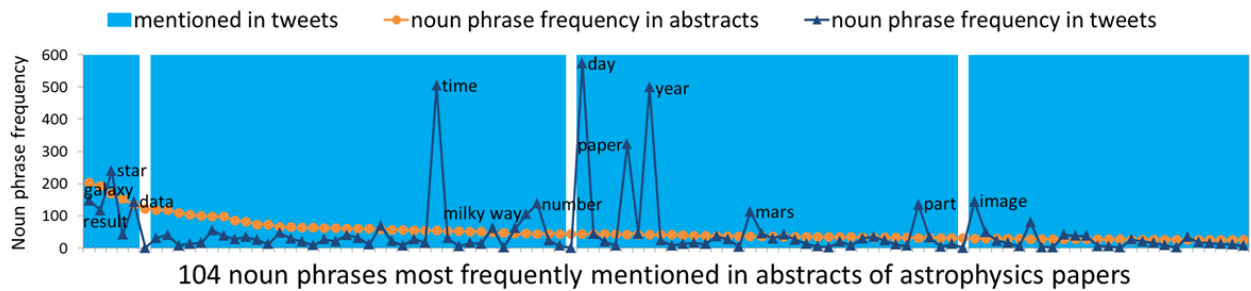

**Figure 2.** Number of appearance of noun phrases in abstracts and tweets by the 18 most frequently publishing astrophysicists for the noun phrases that were mentioned most often (≥25 times) in abstracts. Only three of the 104 most commonly used abstract terms did not appear in tweets.

Interestingly, among the 104 most frequently used terms in abstracts only 3 were not mentioned on Twitter (Figure 2), which points to the fact that the most central (i.e. commonly used) terms from scholarly documents made it to Twitter, whereas less frequent terms did not. A closer analysis of Figure 2 reveals, that those of the most frequent abstract terms mentioned at least 100 times on Twitter are – although astrophysics related – very general terms: *galaxy*, *result*, *star*, *data*, *time*, *milky way*, *number*, *day*, *paper*, *year*, *mars*, *part* and *image*.

On the author level (Table 7), *PFR-TRe1*'s [3] tweets are the most similar to the abstracts of his or her publications. Reflected by a cosine of 0.096, 322 (21.5%) of the noun phrases published in the abstracts also appeared in tweets (Figure 3). Interestingly, this person is also one of the two authors with the highest citation impact, while, on the other hand, the retweet rate, which can be regarded as the Twitter citation rate of the person's tweets, remains far below average. The top five terms in tweets by *PRF-TRe1* were *paper* (179 times), which suggests that parts of his or her tweets refer to scholarly contents, *day* (171), *thank* (147), *python* (101) and *dotastro* [=hashtag used for an astronomy conference] (63). The use of Twitter affordances and URLs remains below the average among the particular values of this group of 18 astrophysicists (and the averages for all 37 astrophysicists shown in Tables 3 to 6).

**Table 7.** Number of documents in WoS published between 2008 and 2012 (*P*), citation rate of *P* (*C/P*), number of tweets collected ($T_{col}$), retweet rate of $T_{col}$ ($RT_{col}/T_{col}$), number of unique noun phrase extracted from tweets ($NP_T$) and abstracts of publications ($NP_P$), overlap ($NP_T \cup NP_P$) and resulting cosine similarities (*cos*) between the two as well as percentages of tweets that are retweets ($T\%_{RT}$), contain usernames ($T\%_@$), hashtags ($T\%_\#$) and URLs ($T\%_{URL}$) for the 18 most frequently publishing astrophysicists. Except for $T_{col}$, data in columns are colored according to the values for each of the indicators from shades of green (maximum value) over white (mean value) to shades of red (minimum) and ordered descendingly by *cos*. Users were made anonymous using group membership as labels (*PFr*: publishing frequently, *PRe*: publishing regularly, *TFr*: tweeting frequently, *TRe*: tweeting regularly, *TOc*: tweeting occasionally, *TRa*: tweeting rarely).

| user | P | C/P | $T_{col}$ | $RT_{col}/T_{col}$ | $NP_T$ | $NP_P$ | $NP_T \cup NP_P$ | cos | $T\%_{RT}$ | $T\%_@$ | $T\%_\#$ | $T\%_{URL}$ |
|---|---|---|---|---|---|---|---|---|---|---|---|---|
| PFr-TRe1 | 46 | 32.8 | 2,832 | 4.5 | 7,491 | 1,501 | 322 | 0.096 | 7.2 | 44.7 | 20.9 | 13.9 |
| PRe-TRe4 | 14 | 4.9 | 2,987 | 4.3 | 5,033 | 540 | 135 | 0.082 | 10.2 | 77.7 | 8.7 | 32.7 |
| PRe-TFr2 | 27 | 11.6 | 323 | 4.1 | 4,809 | 736 | 148 | 0.079 | 4.0 | 67.6 | 9.8 | 23.5 |
| PRe-TOc3 | 29 | 9.8 | 236 | 0.9 | 772 | 992 | 67 | 0.077 | 13.1 | 40.3 | 51.3 | 30.4 |
| PRe-TRe1 | 16 | 18.3 | 3,247 | 55.6 | 5,671 | 548 | 126 | 0.071 | 15.6 | 67.4 | 15.4 | 64.3 |
| PFr-TOc2 | 49 | 32.8 | 1236 | 44.4 | 3,206 | 1,698 | 141 | 0.060 | 46.2 | 72.0 | 42.6 | 94.8 |
| PRe-TOc1 | 33 | 9.3 | 624 | 0.4 | 2,271 | 1,092 | 93 | 0.059 | 4.2 | 28.8 | 54.3 | 67.1 |
| PRe-TOc4 | 22 | 23.2 | 1,302 | 5.2 | 3,189 | 764 | 89 | 0.057 | 11.3 | 65.2 | 53.2 | 29.7 |
| PRe-TFr3 | 19 | 8.1 | 3,256 | 1.1 | 7,011 | 601 | 117 | 0.057 | 7.8 | 48.9 | 52.2 | 18.0 |
| PRe-TFr1 | 37 | 9.6 | 325 | 7.3 | 8,937 | 1,292 | 190 | 0.056 | 23.4 | 83.4 | 47.7 | 17.5 |
| PRe-TRe2 | 15 | 24.3 | 3,214 | 28.7 | 6,958 | 601 | 110 | 0.054 | 15.1 | 69.7 | 32.3 | 26.6 |
| PFr-TRa1 | 48 | 9.4 | 73 | 1.2 | 308 | 1,947 | 41 | 0.053 | 0.0 | 2.7 | 5.5 | 27.9 |
| PFr-TOc1 | 112 | 20.5 | 423 | 150.7 | 1,235 | 1,254 | 62 | 0.050 | 27.2 | 61.5 | 65.0 | 44.9 |
| PRe-TOc2 | 31 | 8.5 | 148 | 9.3 | 539 | 856 | 32 | 0.047 | 29.1 | 53.4 | 38.5 | 14.8 |
| PFr-TOc3 | 48 | 19.1 | 204 | 0.7 | 700 | 167 | 50 | 0.046 | 2.9 | 24.5 | 10.3 | 24.4 |
| PRe-TRe3 | 15 | 30.9 | 2,692 | 25.2 | 5,399 | 508 | 70 | 0.042 | 3.1 | 52.1 | 18.0 | 30.1 |
| PRe-TOc5 | 14 | 19.3 | 502 | 18.8 | 1,252 | 567 | 32 | 0.038 | 14.3 | 80.1 | 20.5 | 71.5 |
| PRe-TFr1 | 22 | 10.9 | 2,002 | 5.8 | 3,697 | 2,489 | 113 | 0.037 | 4.0 | 83.7 | 11.4 | 15.5 |
| total | 597 | 16.9 | 31,458 | 20.5 | 50,854 | 12,970 | 2,075 | 0.081 | 13.3 | 56.9 | 31.0 | 36.0 |

The other astrophysicist with the shared highest citation rate of 32.8 (*PFr-TOc2*) shows the opposite tweeting behavior, as he or she makes extensive use of the Twitter affordances and URLs; almost half of his tweets are retweets or contain hashtags, 72% of them mention at least one user name and 94.8% contain a URL. With an average of 44 retweets per tweet, *PFr-TOc2* has the third highest retweet rate shown in Table 7. Tweets and abstracts are less similar than average (*cos*=0.060) with only 141 noun phrases overlapping between the two sets. The most frequent noun phrases in the abstracts were *active region* (22 times in abstracts; 2 times in tweets), *flux rope* (14; 4) and *eruption* (10; 18) (Figure 4). On Twitter *thank* (243 times), *sun* (237) and *earth* (85) were most often used, of which the latter two also appeared in the abstracts three and four times, respectively.

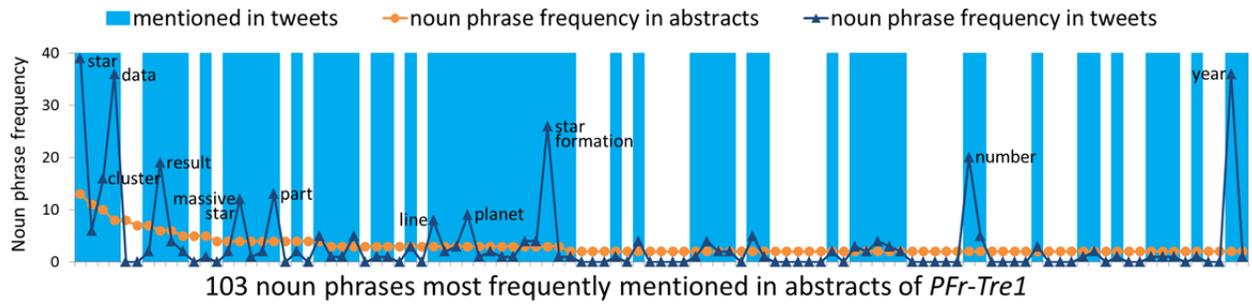

**Figure 3.** Number of appearance of noun phrases in abstracts and tweets by *PFr-Tre1* for 103 noun phrases that were mentioned most often (≥2 times) in abstracts.

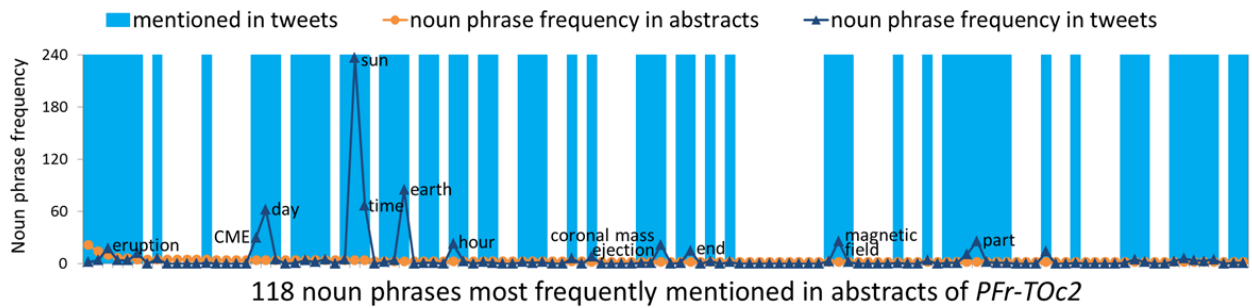

**Figure 4.** Number of appearance of noun phrases in abstracts and tweets by *PFr-TOc2* for 118 noun phrases that were mentioned most often (≥2 times) in abstracts.

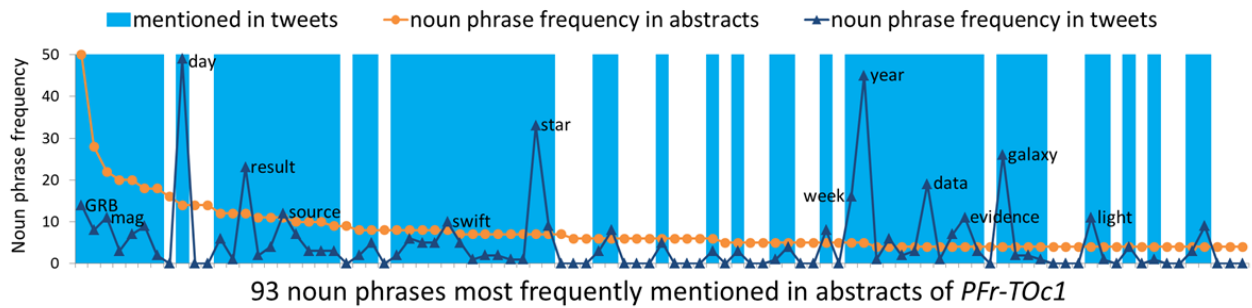

**Figure 5.** Number of appearance of noun phrases in abstracts and tweets by *PFr-TOc1* for 93 noun phrases that were mentioned most often (≥4 times) in abstracts.

The most active researcher (*PFr-TOc1*), as measured by the number of papers published between 2008 and 2012, also makes frequent use of the Twitter affordances and sends more URLs than average. In fact, together with the previously mentioned *PFr-TOc2*, they are the only two among the 18 astrophysicists who consistently retweet and include user names, hashtags and URLs more than average. With a retweet rate of 150.7 (in other words, on average each tweet is forwarded 150 times by others) this is the astrophysicist with the highest impact on Twitter among those analyzed. With a citation rate of 20.5, this person's scholarly publications are also cited more than average. However, the similarity between tweets and abstracts is quite low (*cos*=0.050). Only 62 noun phrases were mentioned both on Twitter and in his or her papers. Of this overlap, *time* (51 times in tweets; 3 times in abstracts), *day* (49; 14), *year* (45; 5), *star* (33; 7) and *point* (28; 1) were most frequently mentioned in tweets, while *GRB* [=gamma-ray burst] (50 times in abstracts; 14 times in tweets), *event* (28; 8), *mag* (22; 11), *burst* (20; 3) and *observation* (20; 7) appeared most often in the abstracts (Figure 5).

Finally, Table 7 shows that no clear trend emerges regarding the similarity between tweets and abstracts and citation and Twitter impact. What can be seen is that the nine astrophysicists who are cited below the average citation rate (among this group), are also all retweeted far less than the average retweet rate of 20.5. On the other hand, a high citation rate does not guarantee high impact on Twitter as shown by *PFr-TRe1*, *PRe-TOc4* and *PFr-TOc3*.

**4 Conclusions and Outlook**

This paper has analyzed the tweeting behavior of 37 selected astrophysicists on Twitter and compared it to their publication behavior and impact based on citations from scholarly documents. While it could be shown that there is a moderate negative correlation between the number of publications and tweets per day, meaning that intensive Twitter users do not publish and that the most active researchers tweet only occasionally, there are astrophysicists that do not fit into this categorization. Different user groups were thus defined according to both tweet and publication frequency, showing that tweeting behavior – such as the use of hashtags, usernames, URLs and sending retweets – differs between them. Frequent Twitter users are more likely to direct tweets to other users by adding @username – which can indicate a stronger need for personal communication or discussion as well as for the building of social networks on Twitter – while frequently publishing authors who tweet at least regularly, occasional authors who tweet rarely, and frequent tweeters who published at least once, are most likely to include URLs, indicating diffusion of information. The comparison of tweet and abstract terms shows that the overlap between the two sets of vocabularies is in general very low, although the more central the term is, based on its appearance in the abstracts of scholarly articles, the more likely it is to be taken up in tweets. Those among the most frequent abstract terms that are also mentioned often on Twitter are, however, very general terms.

On the whole, these results show that astrophysicists who are highly active on social media are less active within the scientific community (as indicated by their number of papers and citation rates), and that the overlap between the topics they discuss on Twitter and those found in scientific abstracts is quite low. Researchers have multidimensional lives – they might be avid runners, foodies, sport fans or enjoy stamp collecting – and their discussions on Twitter might include these various activities (Bauer, 2013). Hence, at least in the case of the astrophysicists analyzed in this paper, researchers' activity on Twitter should not be considered as purely scientific, as Twitter is not restricted to this single type of communication. For those tweets including research-related contents, Twitter provides a fast mean to distribute scientific information to an audience broader than that reading scientific literature. Future research should apply the suggested methods to a larger set of researchers and cover other scientific disciplines to obtain more representative results that will help to distinguish the meaning of tweet counts currently used as a potential indicator of research impact.

**Notes**

[1]   Two users can be considered early adopters as they joined Twitter as early as 2007, while the majority of users (43%) signed up in 2009, nine (24%) joined in 2008, five (14%) in 2010, six (16%) in 2011 and one signed up in April 2012.

[2]   https://apps.facebook.com/selectivetwitter

[3]   Twitter handles were anonymized using group membership as labels, i.e., *PFr*: publishing frequently, *PRe*: publishing regularly, *TFr*: tweeting frequently, *TRe*: tweeting regularly, *TOc*: tweeting occasionally, *TRa*: tweeting rarely.